\documentclass[11pt]{article}

\usepackage[a4paper,margin=1in]{geometry}
\usepackage{amsmath,amssymb}
\usepackage{graphicx}
\usepackage{hyperref}

\title{\bf Bounded cumulative observables from local linear relaxation}

\author{
Sanjeev Kumar Verma\\
Department of Physics and Astrophysics\\
University of Delhi, Delhi 110007, India\\
\texttt{skverma@physics.du.ac.in}
}

\date{}

\begin{document}

\maketitle

\begin{abstract}
Cumulative observables often exhibit saturation in systems involving propagation or spreading with local dissipation. This work shows that bounded cumulative response follows directly from local linear relaxation. Linear cumulative observables accumulated over the lifetime of a relaxing signal are limited by a scale set by the relaxation time, independent of geometry, dimensionality, or microscopic transport dynamics. When relaxation is mapped to space through transport or spreading, this temporal bound yields a corresponding spatial saturation scale determined by the transport law. The result shows that cumulative saturation follows directly from exponential local relaxation and does not depend on the specific transport mechanism.
\end{abstract}


\noindent\textit{1. Introduction.} Signals propagating in physical systems often undergo local relaxation while spreading in time or space. Examples include attenuated waves, diffusion with decay, and stochastic processes with finite memory \cite{chandrasekhar1960,risken1989,gardiner2004}. In many such situations, observables of practical interest are often cumulative rather than local, representing time- or path-integrated response generated by the relaxing signal. Typical examples include optical depth in radiative transfer, accumulated phase in wave propagation, integrated response kernels in linear response theory, and finite correlation time in stochastic dynamics.

Saturation of cumulative response is widely observed: increasing observation time or path length yields progressively smaller incremental contribution. This behaviour is commonly described using system-specific attenuation mechanisms such as scattering statistics, coherence decay, or empirical path-loss relations. In radiative transfer and kinetic theory, attenuation is characterized through optical depth and mean free path \cite{chandrasekhar1960,cercignani1988}. Explanations of saturation are therefore typically formulated within particular physical models rather than as a general consequence of local relaxation itself.

The present work shows that bounded cumulative response follows directly from local linear relaxation. When a signal decays exponentially in time due to a local dissipative mechanism, linear observables defined as time- or path-integrals of the signal approach a finite limit determined by the relaxation time. This boundedness holds independently of geometry, dimensionality, and microscopic transport dynamics. Transport or spreading mechanisms determine only how the relaxation time is mapped onto spatial extent.

More generally, bounded cumulative response follows from integrability of trajectories generated by dissipative linear dynamics. Exponential relaxation ensures that the signal amplitude is integrable over time, producing a finite cumulative response. The same mathematical condition underlies finite correlation time in stochastic processes and convergence of Green--Kubo relations for transport coefficients \cite{vankampen1992,kubo1966}. This separation clarifies two distinct roles: local relaxation determines existence of the saturation bound, while transport determines the spatial scale at which saturation occurs.

Linear relaxation processes arise in many areas of physics, including radiative transfer and kinetic theory \cite{chandrasekhar1960,cercignani1988}, diffusion with decay and stochastic dynamics \cite{risken1989,gardiner2004}, and spectroscopic relaxation phenomena such as nuclear magnetic resonance \cite{abragam1961}. Observables in these systems frequently take cumulative form, including path-integrated attenuation, accumulated phase shifts, integrated response functions, and memory kernels \cite{gardiner2004,vankampen1992}. Across this class of systems, bounded cumulative response provides a simple diagnostic principle: persistent growth of a linear cumulative observable indicates the presence of additional physical ingredients beyond local linear relaxation.

\noindent\textit{2. Minimal framework.} Consider a scalar signal $\psi(t)$ undergoing local linear relaxation with rate $\nu>0$,

\begin{equation}
\frac{d\psi}{dt}=-\nu\psi ,
\end{equation}

and define a linear cumulative observable

\begin{equation}
\mathcal A(T)=\int_0^T \psi(t)\,dt .
\end{equation}

When spatial extent is relevant, distance is related to time through a transport scaling law

\begin{equation}
x\sim \ell(t),
\end{equation}

where $\ell(t)$ denotes the characteristic distance explored by the signal after time $t$, determined by the underlying transport or spreading process. No specific assumptions are made regarding geometry, dimensionality, or microscopic transport dynamics.

\noindent\textit{3. Bounded cumulative response.} Solving the relaxation equation yields

\begin{equation}
\psi(t)=\psi_0 e^{-\nu t},
\end{equation}

so that the cumulative observable becomes

\begin{equation}
\mathcal A(T)
=
\int_0^T \psi_0 e^{-\nu t}\,dt
=
\frac{\psi_0}{\nu}\left(1-e^{-\nu T}\right).
\end{equation}

Hence,

\begin{equation}
\mathcal A(T)\le \frac{\psi_0}{\nu},
\end{equation}

showing that exponential local relaxation guarantees a finite upper bound for linear cumulative observables.

\noindent\textit{Short- and long-time regimes.} The behaviour is governed by the dimensionless parameter

\begin{equation}
\Pi=\nu T,
\end{equation}

which compares the observation time with the relaxation time $\tau=1/\nu$.

For short times $(\Pi\ll1)$, the exponential decay is weak and the cumulative response grows approximately linearly,

\begin{equation}
\mathcal A(T)\simeq \psi_0 T .
\end{equation}

For long times $(\Pi\gg1)$, further contribution to the integral is exponentially suppressed,

\begin{equation}
\mathcal A(T)\simeq \frac{\psi_0}{\nu}.
\end{equation}

The cumulative response therefore transitions from linear growth to saturation once the observation time exceeds the relaxation time.

\noindent\textit{4. Mapping relaxation to spatial accumulation.} Local relaxation defines the intrinsic time scale $\tau=1/\nu$. Spatial limitation of cumulative response follows from the transport relation $x\sim \ell(t)$, which maps the relaxation time onto a characteristic spatial scale.

\noindent\textit{4.1 Ballistic transport.} For propagation with constant speed $v$, the relation $x=vt$ maps the relaxation time $\tau=1/\nu$ onto the characteristic length $L=v/\nu$. The cumulative spatial response becomes

\begin{equation}
\mathcal A_x(L)
=
\psi_0\frac{v}{\nu}\left(1-e^{-\nu L/v}\right),
\end{equation}

which saturates for $L\gg v/\nu$.

\noindent\textit{4.2 Diffusive transport.} For diffusion with decay, spreading obeys $\ell(t)\sim\sqrt{Dt}$, giving the characteristic length $L\sim\sqrt{D/\nu}$. The cumulative spatial response becomes

\begin{equation}
\mathcal A_x(L)
\sim
\psi_0
\frac{\sqrt{\pi D}}{2\sqrt{\nu}}
\mathrm{erf}\!\left(\sqrt{\frac{\nu}{D}}\,L\right),
\end{equation}

which saturates for $L\gg\sqrt{D/\nu}$.

\noindent\textit{4.3 Stochastic relaxation with noise.} For stochastic relaxation,

\begin{equation}
\dot\psi=-\nu\psi+\eta(t),
\end{equation}

with zero-mean noise $\eta(t)$, the mean response decays exponentially, $\langle\psi(t)\rangle=\psi_0 e^{-\nu t}$, consistent with Ornstein--Uhlenbeck relaxation, for which both the mean and the autocorrelation decay exponentially with rate $\nu$ \cite{uhlenbeck1930}. The relaxation rate sets a finite memory time $\tau=1/\nu$ \cite{gardiner2004,vankampen1992}, and integration again yields the finite cumulative value $\psi_0/\nu$.

\noindent\textit{4.4 Effective time--distance mapping.} The preceding cases share the same mathematical structure: linear evolution combined with local exponential relaxation. Transport enters only through the relation between distance and the time required for the signal to reach that distance. Let $x \sim \ell(t)$ denote the characteristic transport law relating distance and time. Inverting this relation defines an effective time $t_{\mathrm{eff}}(x)=\ell^{-1}(x)$, giving

\begin{equation}
\psi(x)=\psi_0 e^{-\nu t_{\mathrm{eff}}(x)}.
\end{equation}

An integral readout along a propagation path of length $L$ becomes

\begin{equation}
\mathcal A_x(L)
=
\psi_0 \int_0^L e^{-\nu t_{\mathrm{eff}}(x)}dx .
\end{equation}

The cumulative response remains finite as $L\to\infty$ if and only if the kernel $e^{-\nu t_{\mathrm{eff}}(x)}$ is integrable over distance,

\[
\int_0^\infty e^{-\nu t_{\mathrm{eff}}(x)}dx < \infty .
\]

Integrability follows directly from exponential local relaxation together with a transport relation for which the effective propagation time increases without bound. Transport therefore determines only the spatial scale over which saturation is approached. Persistent growth of a cumulative observable therefore indicates that the underlying dynamics cannot be described solely by local linear relaxation.

\noindent\textit{5. Discussion.} Exponential relaxation is central to linear response theory, stochastic dynamics, and transport processes, where decay of amplitudes or correlations is characterized by a finite relaxation time \cite{risken1989,gardiner2004,vankampen1992}. In transport theory, exponential attenuation defines characteristic length scales such as mean free paths and optical depths \cite{chandrasekhar1960,cercignani1988}, while stability theory establishes exponential bounds for dissipative linear systems \cite{ogata2010}. From a mathematical perspective, exponential decay ensures integrability of response or correlation kernels. The same integrability condition underlies convergence of Green--Kubo relations for transport coefficients, where finite correlation time guarantees bounded cumulative response \cite{kubo1966,vankampen1992}. In radiative transfer, exponential attenuation of intensity implies convergence of path-integrated deposited energy \cite{chandrasekhar1960}. In stochastic dynamics, exponential decay of correlations produces finite integrated correlation time, as illustrated by Ornstein--Uhlenbeck relaxation \cite{vankampen1992,uhlenbeck1930}. In diffusion with decay, exponential relaxation limits contributions from distances beyond the characteristic length $\sqrt{D/\nu}$, yielding finite spatial accumulation \cite{risken1989}.

Similar behaviour arises in wave propagation through extended media. In radio-frequency propagation, exponential path-loss suppresses contributions from distant propagation segments, so path-integrated response weighted by the decaying signal approaches a finite limit set by the attenuation length \cite{ishimaru1978}. In interferometric systems, cumulative phase fluctuations reduce coherence with increasing baseline length, and characteristic coherence scales quantify distances beyond which additional propagation contributes only weakly to coherent signal formation \cite{fried1966,goodman2015}. In atmospheric wave–optics coupling, wavefront distortions accumulate along extended paths while local attenuation limits the effective contribution of distant segments \cite{tatarskii1961}. These examples illustrate the same structural feature identified above: local attenuation or decorrelation suppresses contributions from distant propagation segments, so cumulative response remains effectively bounded by a scale determined by the underlying relaxation or coherence length.

Exponential local relaxation alone ensures existence of a finite bound for cumulative response, independent of detailed transport-specific assumptions. Transport determines only the spatial scale at which saturation occurs. Conversely, persistent growth of a cumulative observable implies violation of the underlying linear local relaxation structure. Bounded cumulative response thus reflects a general property of dissipative linear dynamics rather than a feature of any specific transport model.

\noindent\textit{Acknowledgement.}  AI-based tools were used to assist with literature navigation and clarity of presentation. All scientific content, derivations, and conclusions were developed and verified solely by the author.

\end{document}